# Parity-Lifted Radiative Degeneracy: Orthogonal Dipoles over Parallel Dipoles in Achiral Dielectric Cavities

Zhong-Jian Yang,* Ya-Pei Wang, Lin Ma, Jian-Hua Liang, Yue You, and Jun He*

*Hunan Key Laboratory of Nanophotonics and Devices, School of Physics, Central South University, Changsha 410083, China*

*zjyang@csu.edu.cn; junhe@csu.edu.cn

**Abstract**:
Parity symmetry enforces radiative degeneracy of parity-conjugated emitters in free space. We selectively lift this degeneracy in an achiral dielectric cavity via coupling-induced global parity breaking: the cavity and individual dipoles each preserve parity, while their fixed relative orientation breaks global parity, enabling differential decay. This mechanism exhibits a structure–function trend in marked contrast to the Rosenfeld rule: orthogonal electric-magnetic dipoles yield radiative asymmetry approaching the theoretical limit of 2, whereas parallel ones show negligible differentiation. A semi-analytical model, validated across multiple modes, confirms generality. These findings establish a new paradigm for symmetry engineering without intrinsic chirality.

Pairs of radiative electric-magnetic dipoles related by spatial inversion (parity transformation)—termed parity-conjugated emitters, and commonly known as enantiomers in molecular systems—exhibit strictly identical total radiative decay rates in free space. This inherent radiative degeneracy is firmly protected by parity symmetry. Spatial inversion maps one parity-conjugated emitter to its counterpart without perturbing the surrounding electromagnetic environment, which enforces equal radiative emission rates universally in free-space systems [1, 2].

Conventional established strategies to break this parity-protected degeneracy rely on constructing intrinsically parity-asymmetric photonic environments [3-13]. Such parity-breaking platforms span non-Hermitian potentials, bi-isotropic chiral nanoparticles, and resonant chiral cavities. Among these platforms with broken inversion symmetry, geometrically chiral cavities and chiral resonant modes represent one of the most prominent geometry-based implementations to introduce predefined environmental asymmetry [9, 13]. A fundamental limitation of these geometry-encoded approaches is that symmetry breaking is encoded in the cavity geometry itself, which inherently limits reconfigurability and often requires complex chiral nanofabrication [14, 15].

Here we present a distinct physical mechanism to achieve efficient differentiation between parity-conjugated emitters within fully achiral, parity-symmetric dielectric cavities. The lifting of radiative degeneracy stems from near-field coupling-induced global parity breaking of the combined emitter–cavity system, rather than intrinsic symmetry breaking of individual components. We observe a counterintuitive structure–property trend in marked contrast to the conventional Rosenfeld rule for angular-resolved chiroptical signals [1, 16]. While the Rosenfeld rule governs the intrinsic rotatory strength of isolated chiral molecules—for which parallel dipoles yield the strongest response—our cavity-based radiative decay mechanism exhibits the opposite tendency: orthogonal electric–magnetic dipoles, which do not discriminate between parity-conjugated emitters in free space, drive the radiative asymmetry factor (defined below) toward the theoretical limit of 2 under ideal single-mode conditions, whereas parallel dipoles preserve parity symmetry and yield negligible differentiation between parity-conjugated emitters. Our work refreshes the fundamental understanding of parity regulation and offers a new paradigm for symmetry engineering and radiative state manipulation in nanophotonics.

Figure 1 illustrates the mechanism of parity symmetry breaking induced by coupling radiative dipoles to an optical cavity mode. In free space, parity symmetry enforces the radiative degeneracy of parity-conjugated emitters: under parity inversion, the electromagnetic environment remains unchanged, so the total radiative decay rates of any parity-conjugated pair—whether in parallel or orthogonal dipole configurations—are strictly equal, independent of the intrinsic chirality of the emitter (Fig. 1a).

Introducing a parity-symmetric, high-index dielectric cavity provides spatially overlapping and orthogonal electric and magnetic near fields via Mie resonances [17-20]. When the electric and magnetic dipoles are parallel, one of the two coupling channels vanishes due to geometric orthogonality, leaving only a single parity-invariant coupling. As a result, the radiative degeneracy is preserved, as illustrated in Fig. 1b. In the orthogonal configuration, both electric and magnetic couplings are simultaneously activated. The vector matching between the dipoles and the cavity fields causes the cross-coupling term to reverse its sign under parity inversion, breaking the effective parity symmetry of the composite system and ultimately lifting the radiative degeneracy (Fig. 1b).

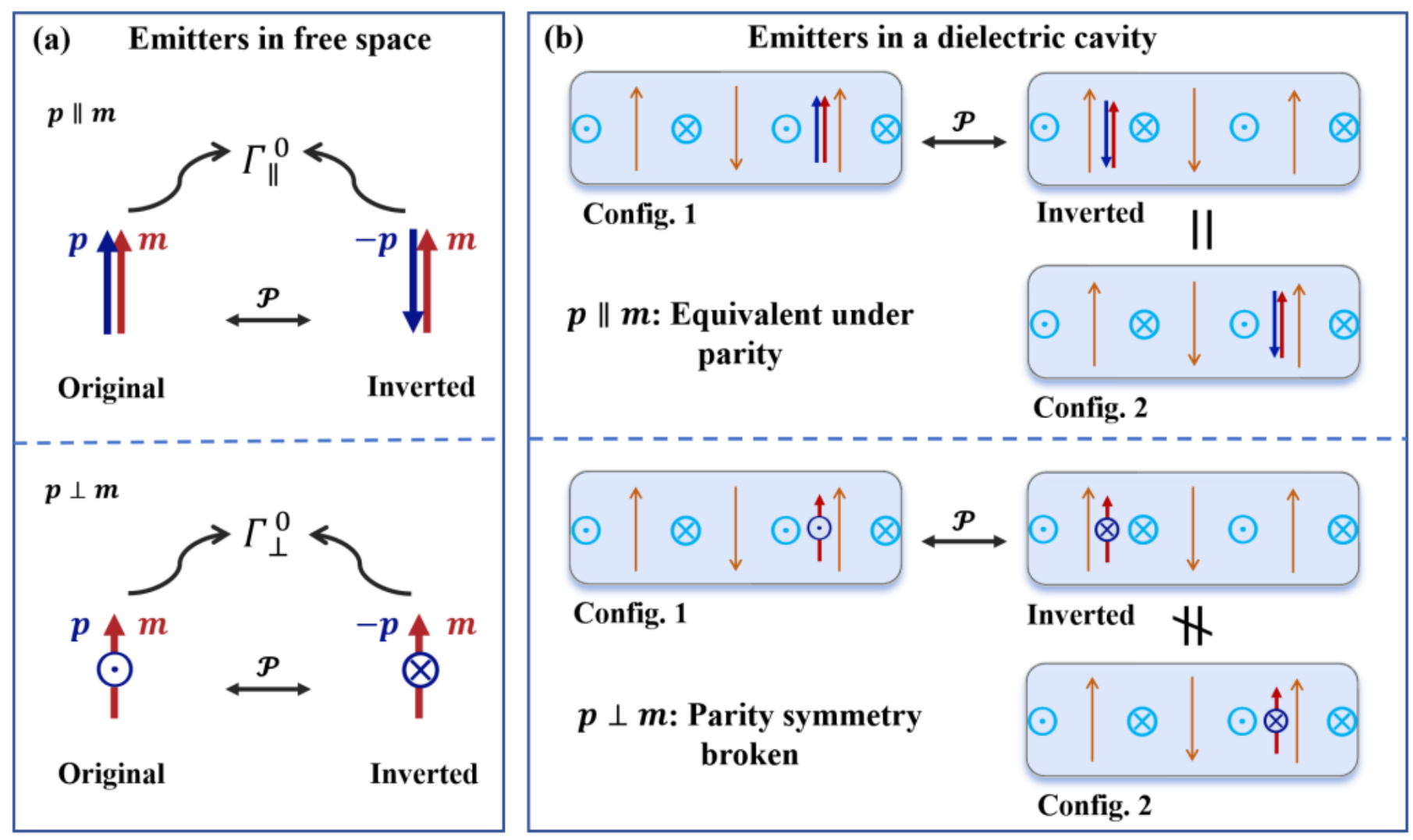

**Figure 1.** Parity behavior of parity-conjugated emitters formed by coupled electric (***p***) and magnetic (***m***) dipoles. (a) Free space: Under parity inversion $\mathcal{P}$, ***p***→−***p***, ***m***→***m***. For both the parallel (***p***//***m***) and orthogonal (***p***⊥***m***) configurations, the original and parity-transformed states have identical radiative decay rates, i.e., the radiative degeneracy is preserved. In the parallel configuration, the two dipoles are slightly offset solely for visual clarity; physically, they are co-located. (b) In an achiral (parity-symmetric) dielectric cavity: For the parallel configuration, the parity-inverted state of Config. 1 is geometrically distinct from Config. 2, yet they remain electromagnetically equivalent under parity inversion, thus preserving global parity symmetry in the radiative response. For the orthogonal configuration, the parity-inverted state of Config. 1 does not correspond to Config. 2, breaking parity and lifting the radiative degeneracy. Arrows: dark blue (***p***), red (***m***), light blue (electric near field), orange (magnetic near field).

To quantitatively validate the coupling-induced parity symmetry breaking mechanism, we utilized a theoretical model for dipole radiation within the Green's function formalism [21, 22] and performed full-wave numerical simulations (see Parts I and II of the SM for details [23]). Let the electric and magnetic dipole moments be $\boldsymbol{p}=X\boldsymbol{p}_0$ and $\boldsymbol{m}=Y\boldsymbol{m}_0$, respectively, where $\boldsymbol{p}_0$ and $\boldsymbol{m}_0$ are normalized reference dipole moments satisfying $|\boldsymbol{m}_0| = c|\boldsymbol{p}_0|$. This ensures their free-space radiative decay rates are equal and normalized to $\Gamma^0_{p_0} = \Gamma^0_{m_0} = 1$. $X$ and $Y$ are dimensionless amplitude scaling factors. In the cavity, the individual radiative decay rates of electric ($\Gamma_{\mathrm{EE}}$) and magnetic ($\Gamma_{\mathrm{MM}}$) dipoles are given by:

$$\Gamma_{\mathrm{EE}} = AX^2, \Gamma_{\mathrm{MM}} = BY^2, \tag{1}$$

where the coefficients $A=\Gamma_{\mathrm{EE}}/\Gamma^0_p$ and $B=\Gamma_{\mathrm{MM}}/\Gamma^0_m$ are precisely the standard electric and magnetic dipole Purcell factors [24, 25], and $\Gamma^0_p = X^2$ and $\Gamma^0_m = Y^2$ are their respective free-space radiative decay rates. $A$ and $B$ are dependent on the resonant properties of the cavity mode and the emitter position relative to the field maxima, and are typically obtained from numerical simulations.

When both electric and magnetic dipoles are placed inside the cavity (Fig. 2a), which is achiral (parity-symmetric) in this work, the magnitudes of the cross-coupling rates satisfy $\Gamma_{\mathrm{EM}} = \Gamma_{\mathrm{ME}}$ according to the Lorentz reciprocity theorem (see Part I in SM for details[23]). For the two parity-conjugated states of a chiral emitter, where the phase difference between ***p*** and ***m*** is ±90 ° [1, 3, 16, 26, 27], the cross-coupling term contributes exclusively to the radiative decay rate (rather than the energy shift). Through systematic calculations on a wide range of dielectric Mie cavity modes, we find that when the electric and magnetic dipoles are orthogonal and aligned with the near-field vectors, the cross-coupling rates obey a geometric mean relation $|\Gamma_{\mathrm{EM}}^{\perp}| = |\Gamma_{\mathrm{ME}}^{\perp}| \approx \sqrt{\Gamma_{\mathrm{EE}}\Gamma_{\mathrm{MM}}}$ in the single-mode dominant regime (see Part III in SM [23]). For a general configuration where the plane spanned by ***p*** and ***m*** is coplanar with the local electric–magnetic near-field plane (Figs. 2a, 2b), and the angle between the dipole polarization directions is $\theta$, this relation generalizes to

$$|\Gamma_{\mathrm{EM}}| = |\Gamma_{\mathrm{ME}}| = \sqrt{\Gamma_{\mathrm{EE}}\Gamma_{\mathrm{MM}}}\sin\theta. \tag{2}$$

Note that $\theta$ is defined as the smallest angle between the two dipoles, restricted to the range $0° \leq \theta \leq 90°$. Values greater than $90°$ are equivalent to $180° - \theta$ and produce identical cross-coupling magnitudes. The cross-coupling term reaches its maximum value for the orthogonal configuration ($\theta = 90°$) and vanishes completely for the parallel configuration ($\theta = 0°$).

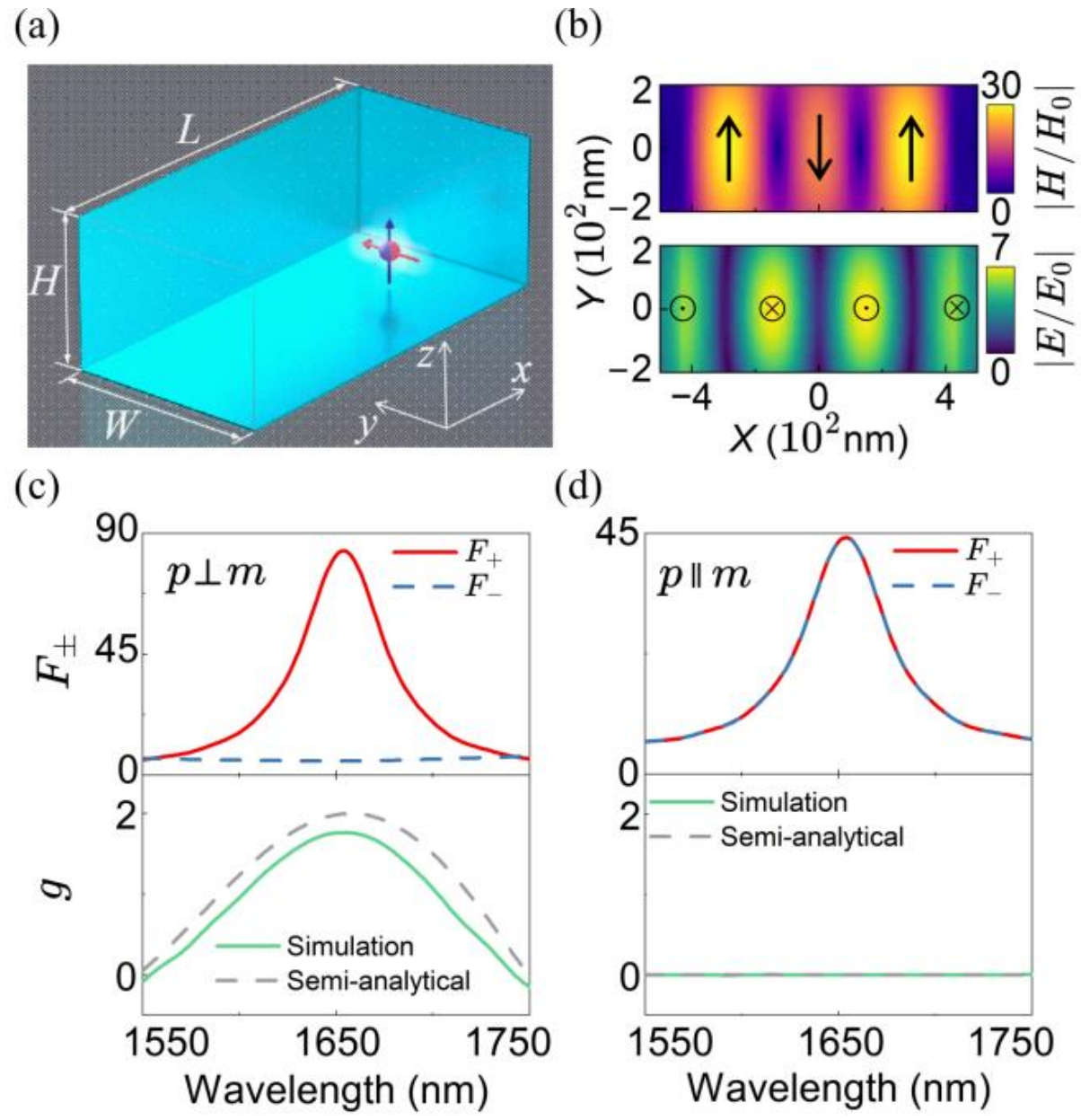

**Figure 2.** Radiative degeneracy lifting in an achiral dielectric cavity. (a) Schematic of orthogonal electric and magnetic dipoles in a dielectric cavity ($n$ = 3.5). The length ($L$), width ($W$) and height ($H$) are 1000, 400 and 400 nm, respectively. (b) Electric and magnetic near fields of the magnetic toroidal mode on the $x$-$y$ plane, with arrows indicating field orientations. (c) Normalized decay rate spectra $F_{\pm}$ for the two parity-conjugated emitters (upper) and radiative asymmetry factor $g$ (lower) for orthogonal dipoles. The emitter's location is $x$ = 250 nm with $y$ = $z$ = 0. (d) Corresponding results for parallel dipoles. Dashed and solid lines for $g$ in (c) and (d) denote semi-analytical and direct simulation results, respectively. In (c) and (d), small square vacuum

voids with a side length of 30 nm are introduced at the emitter positions and their parity-symmetric counterparts.

Combining the phase properties of parity-conjugated states, the relative phase difference between ***p*** and ***m*** is exactly opposite for the two states, leading to opposite signs of the cross-coupling contribution to their total radiative decay rates. This means that for any non-zero cross-coupling, one state will always have a higher total decay rate and the other a lower one. We therefore adopt the convention that $\Gamma_+$ and $\Gamma_-$ denotes the state with the higher and lower total decay rate, respectively. Since $\Gamma_{EM} = \Gamma_{ME}$, the total radiative decay rates can thus be written as $\Gamma_\pm=\Gamma_{EE} + \Gamma_{MM} \pm 2|\Gamma_{EM}|$. Substituting Eqs. (1) and (2) yields:

$$\Gamma_\pm = AX^2 + BY^2 \pm 2XY\sqrt{AB}\sin\theta. \quad (3)$$

We define the radiative asymmetry factor as

$$g = 2\frac{\Gamma_+ - \Gamma_-}{\Gamma_+ + \Gamma_-}, \quad (4)$$

which quantifies the normalized difference in total radiative decay rates between the two parity-conjugated states. This factor is distinct from the conventional circularly polarized luminescence dissymmetry factor that measures the degree of circular polarization of emitted light [1, 28] , and from the differential decay rate between opposite circular polarizations studied in the context of the chiral Purcell effect [27]. Substituting Eq. (3) into Eq. (4), the asymmetry factor for the general coplanar configuration reduces to

$$g = \frac{4XY\sqrt{AB}sin\theta}{AX^2+BY^2}. \quad (5)$$

Eq. (5) holds in the single-mode dominant regime where non-resonant background contributions are negligible. It follows from Eq. (5) that $g$=0 for parallel dipoles ($\theta$=0°), preserving the radiative degeneracy. For orthogonal dipoles ($\theta$=90°) and under the amplitude matching condition $X/Y=\sqrt{B/A}$, the total radiative decay rate of one parity-conjugated state can theoretically vanish, leading to $g$ approaching the fundamental limit of 2.

Full-wave numerical simulations validate the above theoretical predictions (Figs. 2c, d). The electric and magnetic near fields shown in Fig. 2b correspond to the magnetic toroidal mode of a dielectric cavity [20, 29], which provides spatially overlapping and orthogonal field components. The results are presented in terms of the normalized radiative decay rate spectra, $F_{\pm}=\Gamma_{\pm}/(\Gamma^{0}_{p_0}+\Gamma^{0}_{m_0})$. This normalization cancels in Eq. (4), so $g$ is unchanged upon replacing $\Gamma_{\pm}$ with $F_{\pm}$; we therefore use $F_{\pm}$ for all numerical results. Under the amplitude-matching condition $X/Y=\sqrt{B/A}$, for a refractive index of $n$ = 3.5 (corresponding to silicon), the maximal $g$ obtained is $g \approx 1.76$, below the fundamental limit of 2. This deviation mainly originates from the multimode superposition inherent in the dielectric cavity, which hinders the realization of a perfectly isolated single-mode response, together with discretization errors in the numerical simulations. With increasing refractive index, the mode purity improves and $\Gamma_{+}$ increases, driving $g$ progressively toward the fundamental limit of 2 ($g \approx 1.97$ for $n$ = 5; see Fig. S6)[23].

It is worth noting that, to enhance robustness against numerical errors, the emitters are placed inside a small vacuum void embedded in the dielectric cavity in the numerical simulations. This configuration prevents the FDTD solvers from normalizing the radiative decay rates to the dielectric background [25, 30], thereby increasing the apparent numerical values and improving stability against discretization errors (see Part V in SM for more detail [23]). Owing to the high symmetry of these vacuum voids, their introduction does not alter the parity properties of the cavity modes, nor does it affect the core physical mechanism of coupling-induced parity breaking. Unless otherwise noted, all numerical simulations presented in this work are performed with such voids by default. We note that the presence of the voids modifies the specific values of the Purcell factors $A$ and $B$, and consequently shifts the optimal matching condition $X/Y$ relative to the case without voids. This, however, does not affect the theoretical framework or the qualitative conclusions of this work.

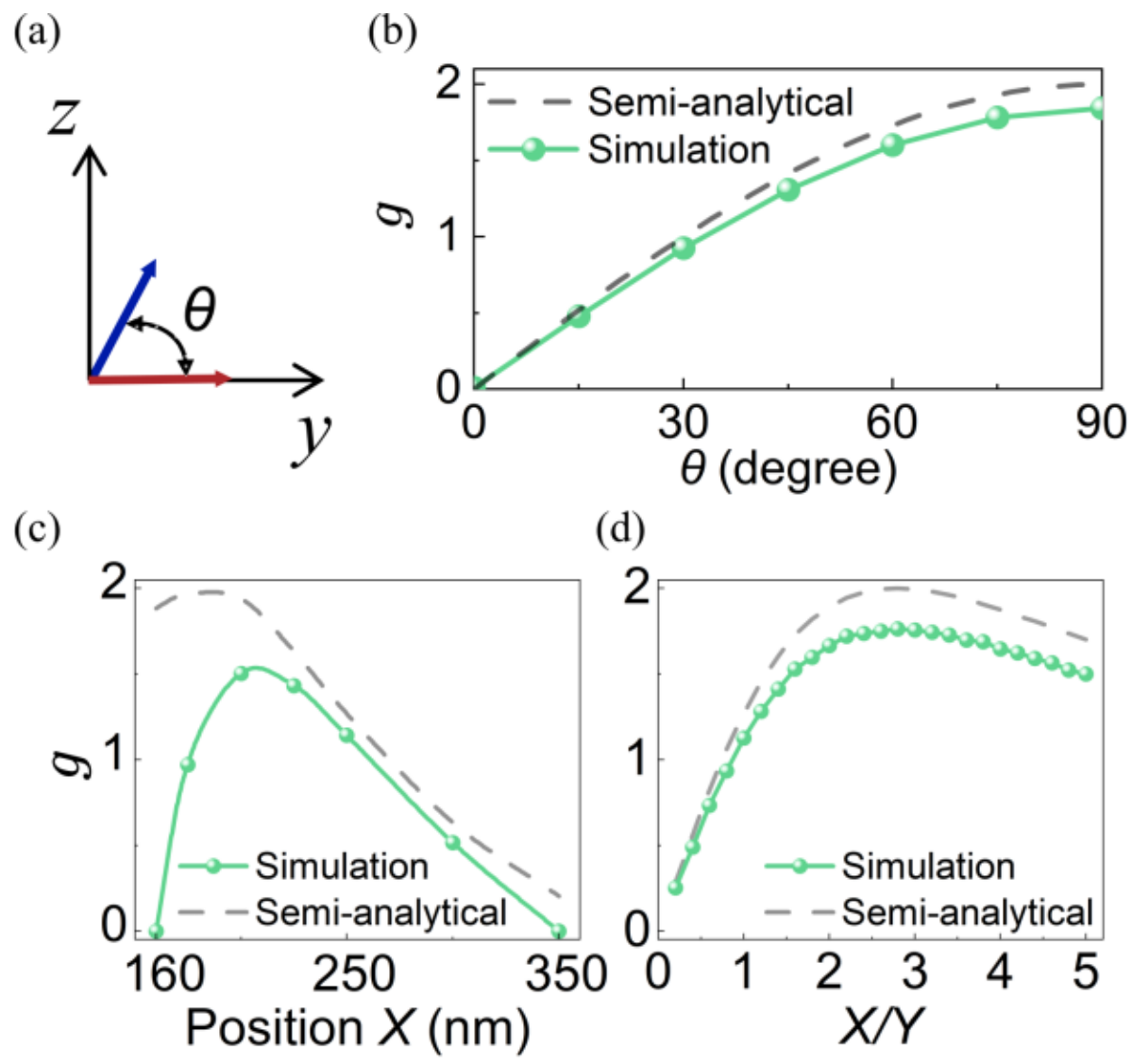

**Figure 3.** Tuning of the radiative asymmetry. (a) Schematic of the dipole angle $\theta$ in the dielectric cavity (other parameters identical to Fig. 2). (b) $g$-factor versus $\theta$. Dashed and solid lines: semi-analytical and simulated results. (c) $g$-factor versus emitter position $x$ at $\theta = 90°$ with a fixed amplitude ratio $X/Y = 1$. (d) $g$-factor versus amplitude ratio $X/Y$ at $\theta = 90°$ with the emitter fixed at $x = 250$ nm with $y = z = 0$.

To further reveal how the dipole configuration governs the radiative asymmetry, we examine the dependence of the $g$-factor on the angle $\theta$ between the electric and magnetic dipoles. For the coplanar geometry defined in Fig. 2, where the plane spanned by $\boldsymbol{p}$ and $\boldsymbol{m}$ is aligned with the local electric–magnetic near-field plane of the cavity, Eq. (5) reduces to $g = 2\sin\theta$ under the amplitude-matching condition $X/Y=\sqrt{B/A}$. This concise relation is validated by full-wave numerical simulations (Figs. 3a and 3b), with deviations between simulation and theory originating from the same factors as discussed for Fig. 2. Notably, this continuous structure–property relation stands in sharp contrast to the conventional Rosenfeld rule [1, 16]: the Rosenfeld rule describes angular-resolved chiroptical signals, predicting the strongest response for parallel dipoles and the weakest for orthogonal ones. In contrast, our radiative decay rate asymmetry factor for parity-conjugated states vanishes for the parallel configuration and reaches its maximum for the orthogonal configuration.

From Eq. (5), for a fixed dipole configuration (i.e., fixed $\theta$), the $g$-factor reaches its upper limit of 2sin$\theta$ predicted by the semi-analytical model when the amplitude ratio satisfies $X/Y=\sqrt{B/A}$ (Figs. 3c and 3d). The ratio $B/A$ can be widely tuned by varying the emitter position within the cavity. At positions where both the electric and magnetic responses are strong—near the maximum of the cross-coupling term $\Gamma_{EM}$ —$B$ is typically larger than $A$, requiring $X > Y$ to satisfy the optimal matching condition (Figs. 3d). This trend aligns naturally with the intrinsic properties of chiral molecules: the magnetic dipole transition moment is generally much weaker than its electric counterpart, and the magnetic Purcell enhancement provided by the cavity effectively compensates for this inherent imbalance. The wide tunability of $B/A$ thus enables matching the optimal emitter position to the intrinsic amplitude ratio $X/Y$ of different chiral molecules, offering substantial flexibility for experimental manipulation of their radiative asymmetry.

We note that the semi-analytical model in Fig. 3c deviates significantly from simulations at positions where the magnetic resonant response vanishes. There, $\Gamma_{MM}$ is dominated by a non-resonant background that does not contribute to the cross-coupling, causing Eq. (2) to overestimate $\Gamma_{EM}$ and yielding an artificially large $g$-factor. After background subtraction, the geometric mean relation holds with high accuracy (see Figs. S5)[23].

The above structure–property relation holds universally across a broad spectral range. Figure 4 extends the analysis to a wide wavelength window covering multiple Mie resonance modes. Here we label the two parity-conjugated states by their fixed relative dipole phases, $F_{+\pi/2}$ and $F_{-\pi/2}$, instead of dynamically sorting by decay rate magnitude. This labeling is fully equivalent to the definition in Eq. (4), with the non-negative $g$ factor given by $g=2|F_{+\pi/2} - F_{-\pi/2}| / (F_{+\pi/2} + F_{-\pi/2})$. For the orthogonal configuration, pronounced non-zero $g$ values appear around all resonant peaks (Fig. 4a). In sharp contrast, the $g$ factor remains near zero across the full spectrum for the parallel configuration (Fig. 4b), confirming the robustness of our conclusion.

For the parallel configuration, the electric–magnetic cross-coupling term vanishes owing to the inherent orthogonality between electric and magnetic fields in dielectric cavity modes. For the orthogonal (or non-parallel) configuration, by contrast, electric and magnetic responses of the cavity mode at the dipole site activate both electric and magnetic coupling channels, breaking the global parity symmetry of the coupled system and yielding a non-zero $g$ factor. This broadband behavior firmly demonstrates that coupling-induced parity breaking is a structural, universal mechanism independent of any specific resonant mode.

The magnitude of the $g$ factor varies across different Mie resonance modes. This quantitative difference is directly rooted in the amplitude matching mechanism revealed in Fig. 3: the extent to which the matching condition $X/Y=\sqrt{B/A}$ is fulfilled differs from mode to mode. This quantitative variation merely reflects differences in the efficiency with which individual modes enhance electric–magnetic cross-coupling. Broadband responses for additional emitter positions confirm the same trend (Fig. S8)[23].

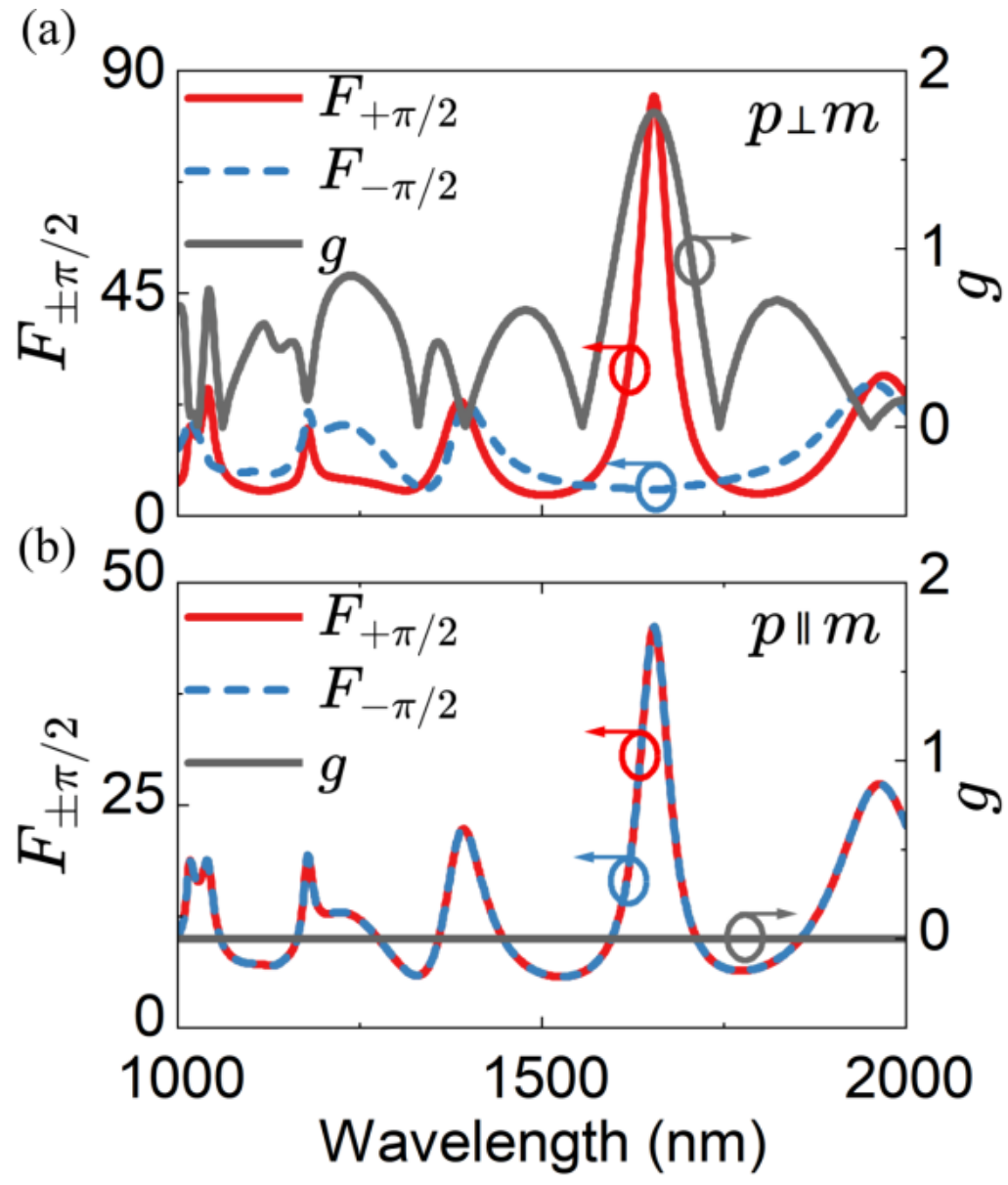

**Figure 4.** Broadband $g$ factor and radiative enhancement for (a) orthogonal ($\theta$=90°) and (b) parallel ($\theta$=0°) dipole configurations. All parameters are identical to those in Fig. 2.

We have demonstrated that the radiative degeneracy between parity-conjugated emitters—protected by parity symmetry in free space—can be selectively lifted in an entirely achiral dielectric cavity via coupling-induced global parity breaking, driven solely by near-field vector matching between the emitter and the cavity mode.

This mechanism exhibits a counterintuitive trend in marked contrast to the conventional structure–function relation described by the canonical Rosenfeld rule: in the cavity, orthogonal dipoles achieve a radiative decay-rate difference factor $g$ approaching the theoretical limit $g = 2$, whereas parallel dipoles—the optimal configuration for conventional chiroptical response—preserve effective parity symmetry and yield negligible decay rate difference.

The generality of this mechanism is corroborated by broadband response, position dependence, and the matching condition between dipole strengths and their respective Purcell factors. The effect relies solely on the relative orientation of dipoles within an achiral cavity—a purely geometric and passive mechanism, independent of conventional chiroptical signals, requiring neither chiral structures nor non-Hermitian engineering. Our findings demonstrate that radiative degeneracy lifting can be achieved in an achiral cavity, opening a new paradigm for symmetry engineering in nanophotonics, where symmetry breaking originates from near-field vector matching instead of intrinsic structural or material chirality.

**ACKNOWLEDGMENTS**
This paper was supported by the National Natural Science Foundation of China (Grant No. 11704416)

—Supplementary material—

# Parity-Lifted Radiative Degeneracy: Orthogonal Dipoles over Parallel Dipoles in Achiral Dielectric Cavities

Zhong-Jian Yang,* Ya-Pei Wang, Lin Ma, Jian-Hua Liang, Yue You, and Jun He*

*Hunan Key Laboratory of Nanophotonics and Devices, School of Physics, Central South University, Changsha 410083, China*

*zjyang@csu.edu.cn; junhe@csu.edu.cn

## Part I: Derivation of the cross-coupling rate

The following derivation assumes a linear, reciprocal dielectric cavity, for which the dyadic Green's function satisfies the Lorentz reciprocity relation.

### 1. Derivation of $\Gamma_{\mathrm{EM}}$ : cross-coupling from a magnetic dipole to an electric dipole

We first compute the electric field $\boldsymbol{E_M}$ radiated by a magnetic dipole located at $\boldsymbol{r_2}$, The equivalent current density of the magnetic dipole is

$$\boldsymbol{J_M}(\boldsymbol{r}) = \nabla \times [\boldsymbol{m}\delta(\boldsymbol{r} - \boldsymbol{r_2})]. \tag{S1}$$

For an arbitrary current source, the electric field is given by the electric dyadic Green's function $\overline{\overline{\boldsymbol{G}}}_{EE}$ as

$$\boldsymbol{E_M}(\boldsymbol{r}) = i\omega\mu_0 \int \overline{\overline{\boldsymbol{G}}}_{EE}(\boldsymbol{r}, \boldsymbol{r}') \cdot \boldsymbol{J_M}(\boldsymbol{r}')\, \mathrm{d}^3\boldsymbol{r}'. \tag{S2}$$

Substituting the magnetic source $\boldsymbol{J_M}$ yields

$$\boldsymbol{E_M}(\boldsymbol{r}) = i\omega\mu_0 \int \overline{\overline{\boldsymbol{G}}}_{EE}(\boldsymbol{r}, \boldsymbol{r}') \cdot (\nabla' \times [\boldsymbol{m}\delta(\boldsymbol{r}' - \boldsymbol{r_2})])\, \mathrm{d}^3\boldsymbol{r}'. \tag{S3}$$

Using the vector identity for integration by parts,

$$\int_V \overline{\overline{\boldsymbol{A}}} \cdot (\nabla' \times \boldsymbol{B}) d^3\boldsymbol{r}' = \int_V \left(\nabla' \times \overline{\overline{\boldsymbol{A}}}\right) \cdot \boldsymbol{B} d^3\boldsymbol{r}'. \tag{S4}$$

The integral can be reduced exactly to the well-known electric–magnetic Green's function relation:

$$\boldsymbol{E_M}(\boldsymbol{r}) = i\omega\mu_0 \left[\nabla \times \overline{\overline{\boldsymbol{G}}}_{EE}(\boldsymbol{r}, \boldsymbol{r_2})\right] \cdot \boldsymbol{m}. \tag{S5}$$

This is the electric field radiated by the magnetic dipole at $\boldsymbol{r_2}$, expressed in terms of the electric Green's function. It forms the basis for evaluating the cross-coupling rate $\boldsymbol{\Gamma_{EM}}$ experienced by an electric dipole.

The electric dipole source $\boldsymbol{J_E}$ is placed in the electric field $\boldsymbol{E_M}$ radiated by the magnetic dipole. Its classical dissipated power is

$$\boldsymbol{P_{E\leftarrow M}} = \frac{1}{2} Re\{\int \boldsymbol{J_E^*}(\boldsymbol{r}) \cdot \boldsymbol{E_M}(\boldsymbol{r}) d^3\boldsymbol{r}\}. \tag{S6}$$

Substituting $\boldsymbol{J_E^*}(\boldsymbol{r}) = i\omega \boldsymbol{p}^* \delta(\boldsymbol{r} - \boldsymbol{r_2})$ and using the sifting property of the $\delta$ function yields

$$\boldsymbol{P_{E\leftarrow M}} = \frac{1}{2} \mathrm{Re}\{(i\omega \boldsymbol{p}^*) \cdot \boldsymbol{E_M}(\boldsymbol{r_2})\}. \tag{S7}$$

Inserting $\boldsymbol{E_M}(\boldsymbol{r_2})$ obtained before gives

$$\boldsymbol{P_{E\leftarrow M}} = -\frac{1}{2} \omega^2 \mu_0 \mathrm{Re}\left\{\boldsymbol{p}^* \cdot \left[\nabla \times \overline{\overline{\boldsymbol{G}}}_{EE}(\boldsymbol{r}, \boldsymbol{r_2})\right]_{\boldsymbol{r}=\boldsymbol{r_1}} \cdot \boldsymbol{m}\right\}. \tag{S8}$$

According to the standard correspondence between quantum mechanics and classical electrodynamics, the spontaneous emission rate is related to the classical dissipated power by $\Gamma = \frac{4P}{\hbar\omega}$. Substituting $\boldsymbol{P_{E\leftarrow M}}$ from the previous step gives

$$\Gamma_{EM} = \frac{4}{\hbar\omega}\left(-\frac{1}{2}\omega^2 \mu_0 \mathrm{Re}\{\boldsymbol{p}^* \cdot \left[\nabla \times \overline{\overline{\boldsymbol{G}}}_{EE}\right] \cdot \boldsymbol{m}\}\right) = -\frac{2\omega\mu_0}{\hbar} \mathrm{Re}\left\{\boldsymbol{p}^* \cdot \left[\nabla \times \overline{\overline{\boldsymbol{G}}}_{EE}\right] \cdot \boldsymbol{m}\right\}. \tag{S9}$$

Using the vacuum light-speed relation $\mu_0 = \frac{1}{\varepsilon_0 c^2}$, the cross-coupling spontaneous emission rate from the magnetic dipole to the electric dipole is finally obtained as

$$\Gamma_{EM} = -\frac{2\omega}{\hbar\varepsilon_0 c^2} \mathrm{Re}\left\{\boldsymbol{p}^* \cdot \left[\nabla \times \overline{\overline{\boldsymbol{G}}}_{EE}(\boldsymbol{r}, \boldsymbol{r_2})\right]_{\boldsymbol{r}=\boldsymbol{r_1}} \cdot \boldsymbol{m}\right\}. \tag{S10}$$

This expression represents the enhancement of the spontaneous emission of an electric dipole at $\boldsymbol{r_2}$ induced by a magnetic dipole at $\boldsymbol{r} = \boldsymbol{r_1}$, where the curl operator $\nabla$ acts on the second spatial variable $\boldsymbol{r_2}$. For co-located dipoles ($\boldsymbol{r_1} = \boldsymbol{r_2}$), this formula reduces to the cross-coupling rate $\Gamma_{EM}$ used in the main text.

**2. Derivation of $\Gamma_{ME}$ : cross-coupling from an electric dipole to a magnetic dipole**

We first compute the electric field $\boldsymbol{E_E}(\boldsymbol{r})$ radiated by an electric dipole located at $\boldsymbol{r_1}$. Under the time-harmonic convention $e^{-i\omega t}$, the equivalent current density of the

electric dipole is

$$\boldsymbol{J_E}(\boldsymbol{r}) = -i\omega\boldsymbol{p}\boldsymbol{\delta}(\boldsymbol{r}-\boldsymbol{r_1}). \tag{S11}$$

For an arbitrary current source, the radiated electric field is given by the electric dyadic Green's function $\overline{\overline{\boldsymbol{G}}}_{\boldsymbol{EE}}$ as

$$\boldsymbol{E_E}(\boldsymbol{r}) = i\omega\mu_0 \int \overline{\overline{\boldsymbol{G}}}_{\boldsymbol{EE}}(\boldsymbol{r},\boldsymbol{r}') \cdot \boldsymbol{J_E}(\boldsymbol{r}')\,\mathbf{d}^{\mathbf{3}}\boldsymbol{r}'. \tag{S12}$$

Substituting $\boldsymbol{J_E}(\boldsymbol{r}')$ and using the sifting property of the $\boldsymbol{\delta}$ function, one obtains

$$\boldsymbol{E_E}(\boldsymbol{r}) = i\omega\mu_0 \cdot (-i\omega)\overline{\overline{\boldsymbol{G}}}_{\boldsymbol{EE}}(\boldsymbol{r},\boldsymbol{r_1}) \cdot \boldsymbol{p} = \omega^2\mu_0\overline{\overline{\boldsymbol{G}}}_{\boldsymbol{EE}}(\boldsymbol{r},\boldsymbol{r_1}) \cdot \boldsymbol{p}. \tag{S13}$$

This is the electric field generated by the electric dipole throughout space, expressed in terms of the electric Green's function. It serves as the starting point for evaluating the cross-coupling rate $\Gamma_{\mathbf{ME}}$ experienced by a magnetic dipole.

The magnetic dipole interacts only with the magnetic field. Using Faraday's law, the magnetic field is obtained from the electric field derived in above.Under the time-harmonic convention, $\nabla \times \boldsymbol{E} = -\frac{\partial \boldsymbol{B}}{\partial \boldsymbol{t}} = i\omega\boldsymbol{B}$, giving

$$\boldsymbol{B}(\boldsymbol{r}) = \frac{1}{i\omega}\nabla \times \boldsymbol{E_E}(\boldsymbol{r}) = -\frac{i}{\omega}\nabla \times \boldsymbol{E_E}(\boldsymbol{r}). \tag{S14}$$

Substituting $\boldsymbol{E_E}(\boldsymbol{r})$ above and evaluating at $\boldsymbol{r} = \boldsymbol{r_2}$ yields

$$\boldsymbol{B}(\boldsymbol{r_2}) = -i\omega\mu_0 \left[\nabla \times \overline{\overline{\boldsymbol{G}}}_{\boldsymbol{EE}}(\boldsymbol{r},\boldsymbol{r_1})\right]_{\boldsymbol{r}=\boldsymbol{r_2}} \cdot \boldsymbol{p}. \tag{S15}$$

This is the magnetic field generated by the electric dipole at $\boldsymbol{r_2}$, serving as the key intermediate result for evaluating the cross-coupling rate $\Gamma_{\mathbf{ME}}$.

The magnetic dipole $\boldsymbol{m}$ located at $\boldsymbol{r_2}$ interacts with the magnetic field $\boldsymbol{B}(\boldsymbol{r_2})$ derived above. Its classical dissipated power is

$$\boldsymbol{P_{M\leftarrow E}} = \frac{1}{2}\omega\mathbf{Re}\{i\boldsymbol{m}^* \cdot \boldsymbol{B}(\boldsymbol{r_2})\}. \tag{S16}$$

Substituting the expression for $\boldsymbol{B}(\boldsymbol{r_2})$ gives

$$\boldsymbol{P_{M\leftarrow E}} = \frac{1}{2}\omega^2\mu_0\mathrm{Re}\left\{\boldsymbol{m}^* \cdot \left[\nabla \times \overline{\overline{\boldsymbol{G}}}_{\boldsymbol{EE}}(\boldsymbol{r},\boldsymbol{r_1})\right]_{\boldsymbol{r}=\boldsymbol{r_2}} \cdot \boldsymbol{p}\right\}. \tag{S17}$$

Using the quantum-classical correspondence $\boldsymbol{\Gamma} = \frac{\mathbf{4}}{\hbar\boldsymbol{\omega}}\boldsymbol{P_{class}}$, the spontaneous emission rate is obtained as

$$\Gamma_{\mathrm{ME}} = \frac{4}{\hbar\omega}\left(\frac{1}{2}\omega^2\mu_0\mathrm{Re}\{\boldsymbol{m}^* \cdot \left[\nabla \times \overline{\overline{\boldsymbol{G}}}_{\boldsymbol{EE}}(\boldsymbol{r},\boldsymbol{r_1})\right]_{\boldsymbol{r}=\boldsymbol{r_2}} \cdot \boldsymbol{p}\}\right)$$

$$= \frac{2\omega\mu_0}{\hbar} \mathrm{Re}\{\boldsymbol{m}^* \cdot \left[\nabla \times \overline{\overline{\boldsymbol{G}}}_{\boldsymbol{EE}}(\boldsymbol{r}, \boldsymbol{r_1})\right]_{\boldsymbol{r}=\boldsymbol{r_2}} \cdot \boldsymbol{p}\}. \quad \text{(S18)}$$

Inserting the vacuum light-speed relation $\mu_0 = \frac{1}{\varepsilon_0 c^2}$ yields the final expression for the cross-coupling spontaneous emission rate from the electric dipole to the magnetic dipole:

$$\Gamma_{\mathrm{ME}} = \frac{2\omega}{\hbar\varepsilon_0 c^2} \mathrm{Re}\left\{\boldsymbol{m}^* \cdot \left[\nabla \times \overline{\overline{\boldsymbol{G}}}_{\boldsymbol{EE}}(\boldsymbol{r}, \boldsymbol{r_1})\right]_{\boldsymbol{r}=\boldsymbol{r_2}} \cdot \boldsymbol{p}\right\}. \quad \text{(S19)}$$

The curl operator $\nabla$ acts on the first spatial variable $\boldsymbol{r_1}$. For co-located dipoles ($\boldsymbol{r_1} = \boldsymbol{r_2}$), this formula reduces to the cross-coupling rate $\boldsymbol{\Gamma}_{\mathrm{ME}}$ used in the main text.

### 3. Reciprocity relation between the cross-coupling rates

Using the Lorentz reciprocity relation for the dyadic Green's function,

$$\boldsymbol{\nabla_2} \times \overline{\overline{\boldsymbol{G}}}_{\boldsymbol{EE}}(\boldsymbol{r_1}, \boldsymbol{r_2}) = -\left[\boldsymbol{\nabla_1} \times \overline{\overline{\boldsymbol{G}}}_{\boldsymbol{EE}}(\boldsymbol{r_2}, \boldsymbol{r_1})\right]^T, \quad \text{(S20)}$$

one obtains $\boldsymbol{p}*\cdot[\nabla\times\mathbf{G}]\cdot\boldsymbol{m}=-[\boldsymbol{m}*\cdot[\nabla\times\mathbf{G}]\cdot\boldsymbol{p}]*$, Since $\Gamma_{\mathrm{EM}}$ and $\Gamma_{\mathrm{ME}}$ are respectively proportional to the real parts of the two sides of this equality, it follows immediately that $\Gamma_{\mathrm{EM}} = \Gamma_{\mathrm{ME}}$. This relation confirms that the two cross-coupling rates are strictly equal in magnitude. Their relative sign is determined by the dipole phase relation, which is why the main text uses the absolute value $|\Gamma_{\mathrm{EM}}|$ in the total decay rate expressions.

## Part II: FDTD methods

The numerical calculations are performed using a commercial finite-difference time-domain (FDTD) solver (Ansys Lumerical). The electric and magnetic dipole emitters are modeled as point sources, and their relative orientation angle $\theta$ can be varied to cover the full range from 0 ° to 90 °. To avoid the default background-medium normalization of radiative decay rates implemented in FDTD solvers, the dipole sources are placed inside small vacuum voids (side length 30 nm) embedded in the dielectric cavity by default (see the discussion on the void configuration in Part V). A uniform mesh size of $\Delta x = \Delta y = \Delta z = 2$ nm is applied in the region surrounding the emitters to ensure sufficient spatial resolution. The dielectric cavity has a refractive index of $n$ = 3.5 (corresponding to silicon), and the surrounding medium is set to $n$ = 1 (vacuum). Perfectly matched layer (PML) boundary conditions are imposed in all three spatial directions. The radiative decay rates are extracted from the total power radiated by the dipoles, and the cross-coupling rates are obtained by subtracting the radiative rates of the individual electric and magnetic dipoles from the combined emitter configuration. All simulations are performed in the time domain with a broadband excitation, and the spectral responses are obtained via Fourier transform.

## Part III: Validation of the geometric mean relation

The geometric mean approximation (Eq. (2) in the main text) serves as the core building block of our semi-analytical model and is systematically validated across multiple resonant mode types. Figures S1-S3 present the validation for the magnetic toroidal mode, magnetic dipole mode, and a higher-order electromagnetic mode. The emitter has an orthogonal configuration, and $|\Gamma_{EM}|$ is denoted as $|\Gamma_{EM}^{\perp}|$ in the main text. Each figure includes a schematic of the cavity structure with field distributions and emitter positions, along with a scatter plot quantifying the correlation between the directly extracted $|\Gamma_{EM}|$ and the geometric mean prediction. Figure S1 additionally provides representative spectra used to extract peak decay rates.

Across all three mode types, the geometric mean relation holds with comparable accuracy when the raw total decay rates are used. After isolating the pure resonant components via background subtraction, the agreement is significantly improved in all cases. This observation confirms that the cross-coupling is dominated by the resonant cavity mode, while off-resonant higher-order multipole backgrounds contribute negligibly to the cross term.

Notably, for the magnetic toroidal mode, the deviations between the raw total-rate prediction and the directly extracted $\Gamma_{EM}$ are most pronounced at positions where the magnetic resonant response approaches zero. There, the magnetic decay rate $\Gamma_{MM}$ is dominated by the non-resonant background, which inflates the geometric mean prediction of $\Gamma_{EE}$ and $\Gamma_{MM}$ and leads to an artificially large $\Gamma_{EM}$. In contrast, at positions where the resonant cross-coupling is strong, the background influence is negligible even without subtraction. For the magnetic dipole mode, the overall deviations are larger because the magnetic resonant response itself is weaker, making the background contribution relatively more significant throughout the spatial range.

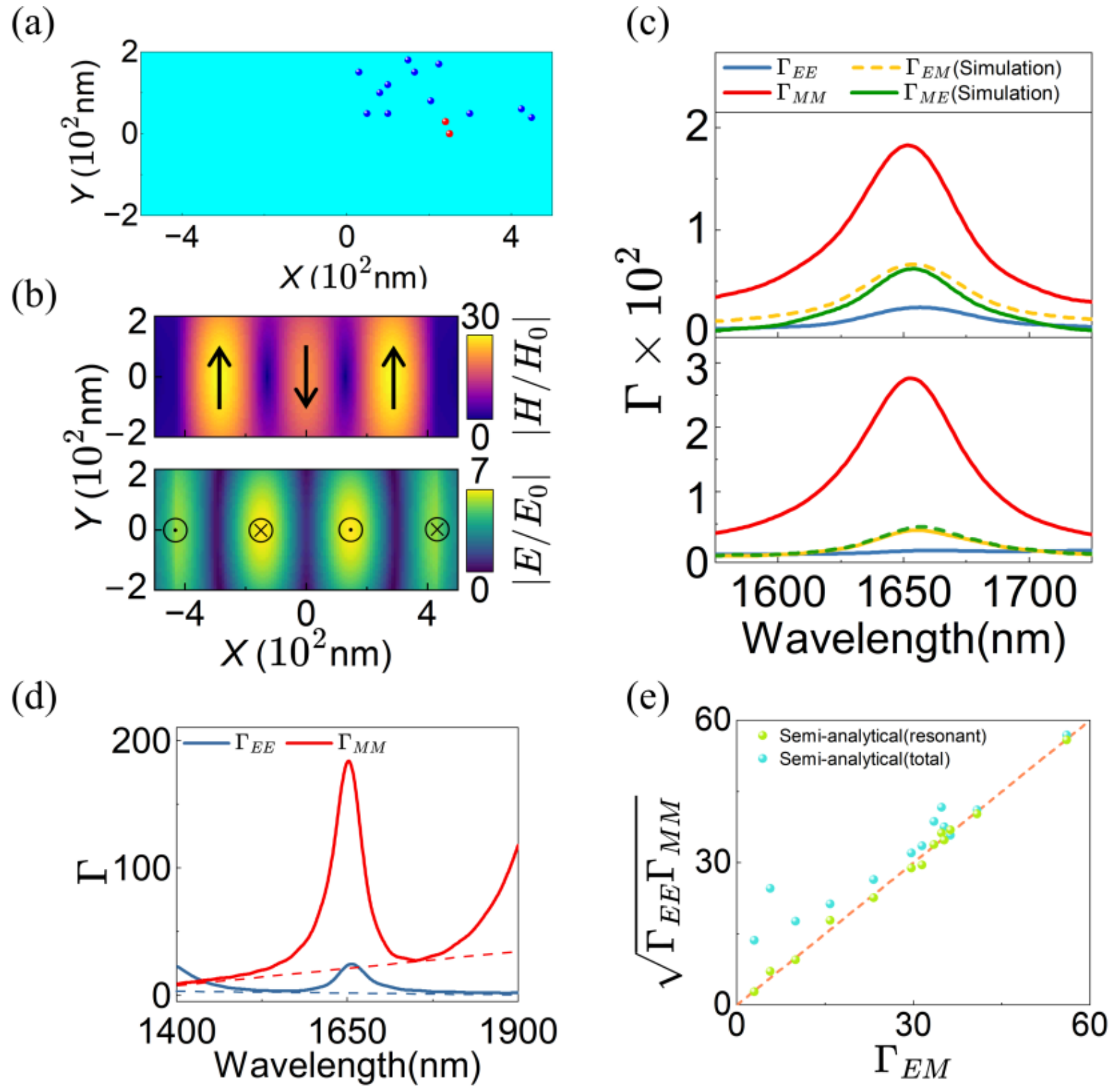

**Figure S1.** Validation of the geometric mean relation for the magnetic toroidal mode. (a) Schematic of the cavity structure with field distribution and emitter positions. The structure is the same as that in Fig. 2 in the main text. (b) Representative decay rate spectra for the orthogonal configuration ($\theta$ = 90°) directly obtained from simulations: $\Gamma_{EE}$, $\Gamma_{MM}$, and the cross-coupling rates $\Gamma_{EM}$ and $\Gamma_{ME}$, plotted as functions of wavelength. The emitter locations are marked as red dots in (a). (c) Illustration of the background subtraction procedure: the non-resonant background level of each spectrum is determined from the off-resonance spectral region (dashed lines), and subtracted from the raw total single-dipole rate to obtain the resonant component. (d) Geometric mean prediction $\sqrt{\Gamma_{EE}\Gamma_{MM}}$ versus the directly extracted $|\Gamma_{EM}|$ at resonances. Blue dots: raw total single-dipole rates including off-resonant background. Green dots: resonant components after background subtraction. The dashed line indicates $y = x$, namely $|\Gamma_{EM}| = \sqrt{\Gamma_{EE}\Gamma_{MM}}$. Data points are collected from multiple emitter positions as shown in (a).

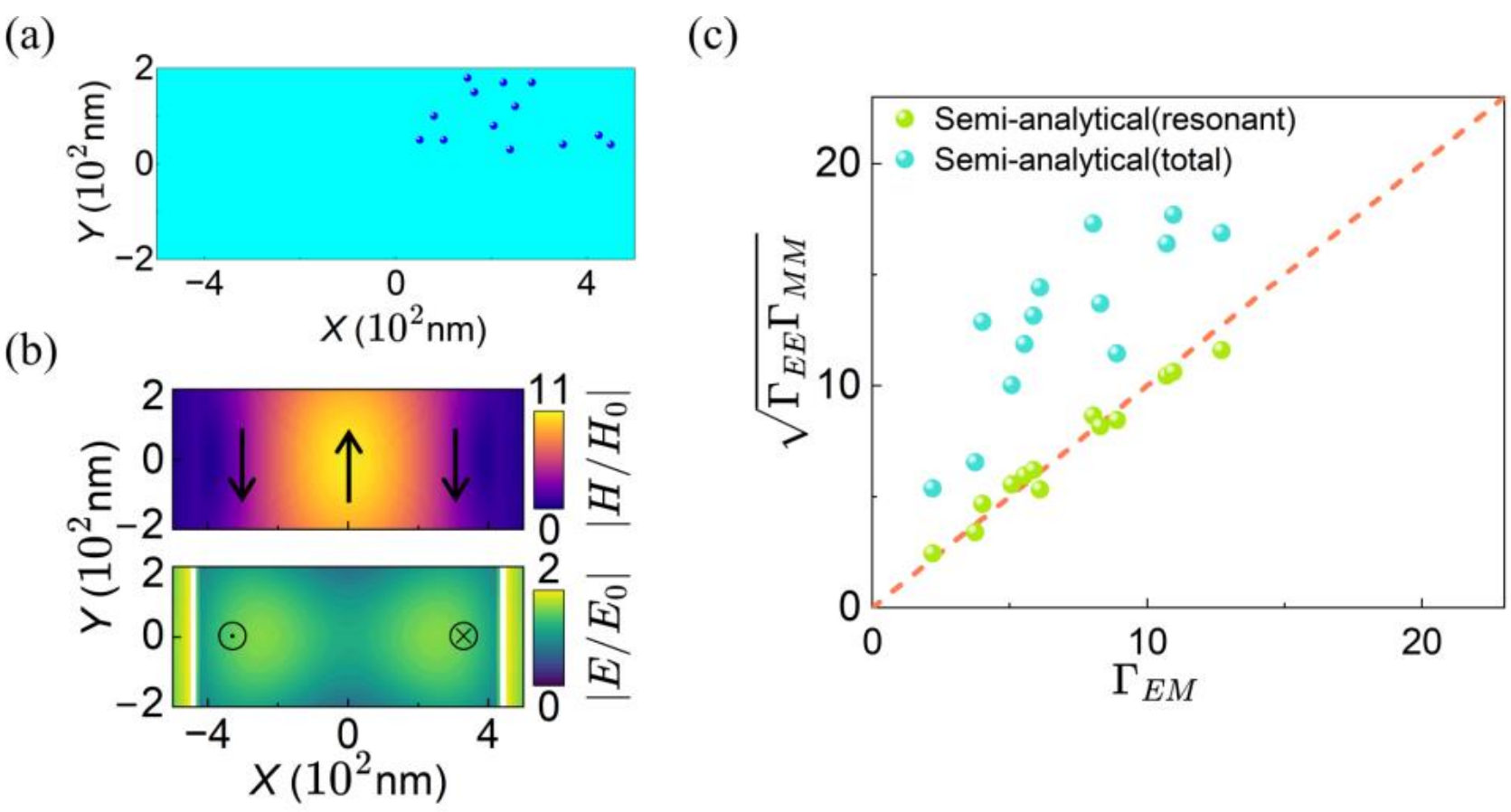

Figure S2. Validation of the geometric mean relation for a magnetic dipole mode. (a) Schematic of the cavity structure with field distributions and emitter positions. The structure is the same as that in Fig. S1. (b) Same as Fig. S1(c), but for the magnetic dipole mode. The background subtraction procedure follows the same method described in Fig. S1(c).

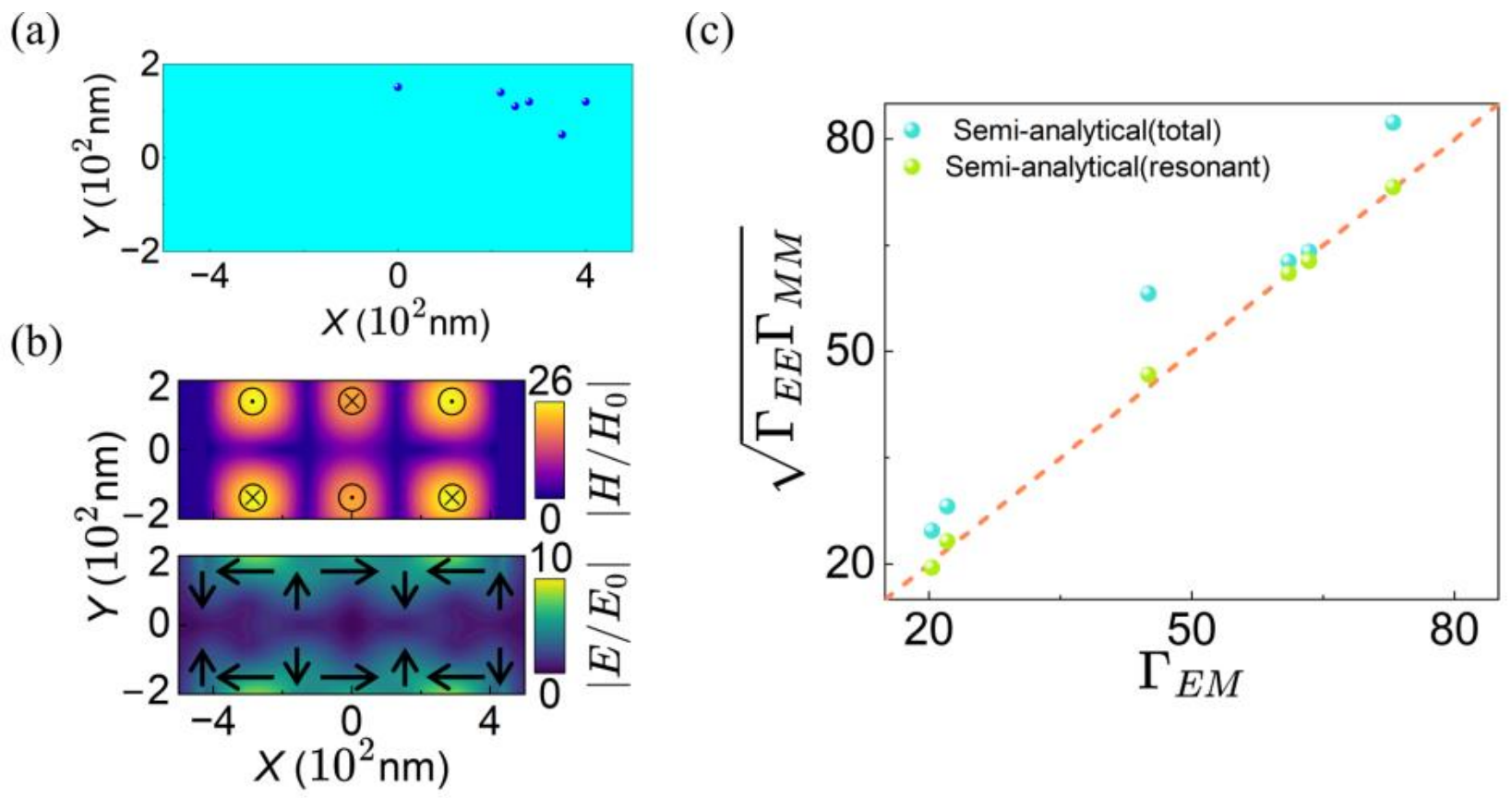

Figure S3. Validation of the geometric mean relation for a higher-order electromagnetic mode of the dielectric cavity. (a) Schematic of the cavity structure with field distributions and emitter positions. The structure is the same as that in Figs. S1 and S2. The mode features a more complex near-field distribution with six magnetic field lobes. (b) Same as Fig. S2(c), but for the higher-order electromagnetic mode. The background subtraction procedure follows the same method described in Fig. S1(c).

Figure S4 further validates the angular dependence $|\Gamma_{\mathrm{EM}}| \propto \sin\theta$ predicted by Eq. (2) for the magnetic toroidal mode at a representative emitter position. At this position, where the resonant cross-coupling is strong, the predictions based on the total and background-subtracted rates are nearly indistinguishable.

This physically motivated single-mode approximation forms the foundation of the semi-analytical model presented in the main text.

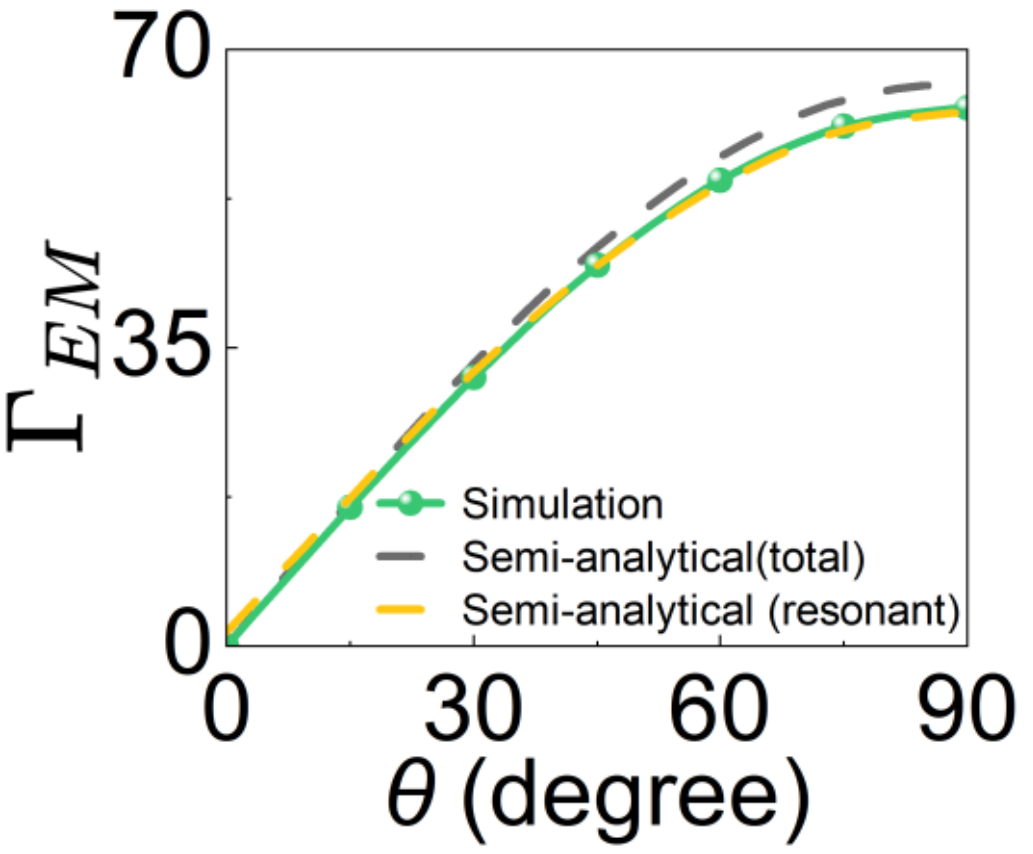

**Figure S4.** Validation of the angular dependence in the geometric mean relation for the magnetic toroidal mode at $x$=250 nm ($y$ = $z$ = 0). Directly simulated $|\Gamma_{\mathrm{EM}}|$ plotted against the angle $\theta$, compared with the geometric mean predictions $\sqrt{\Gamma_{\mathrm{EE}}\Gamma_{\mathrm{MM}}}\sin\theta$ based on the total single-dipole rates (gray line, "Semi-analytical (total)") and the resonant components after background subtraction (yellow line, "Semi-analytical (resonant)").

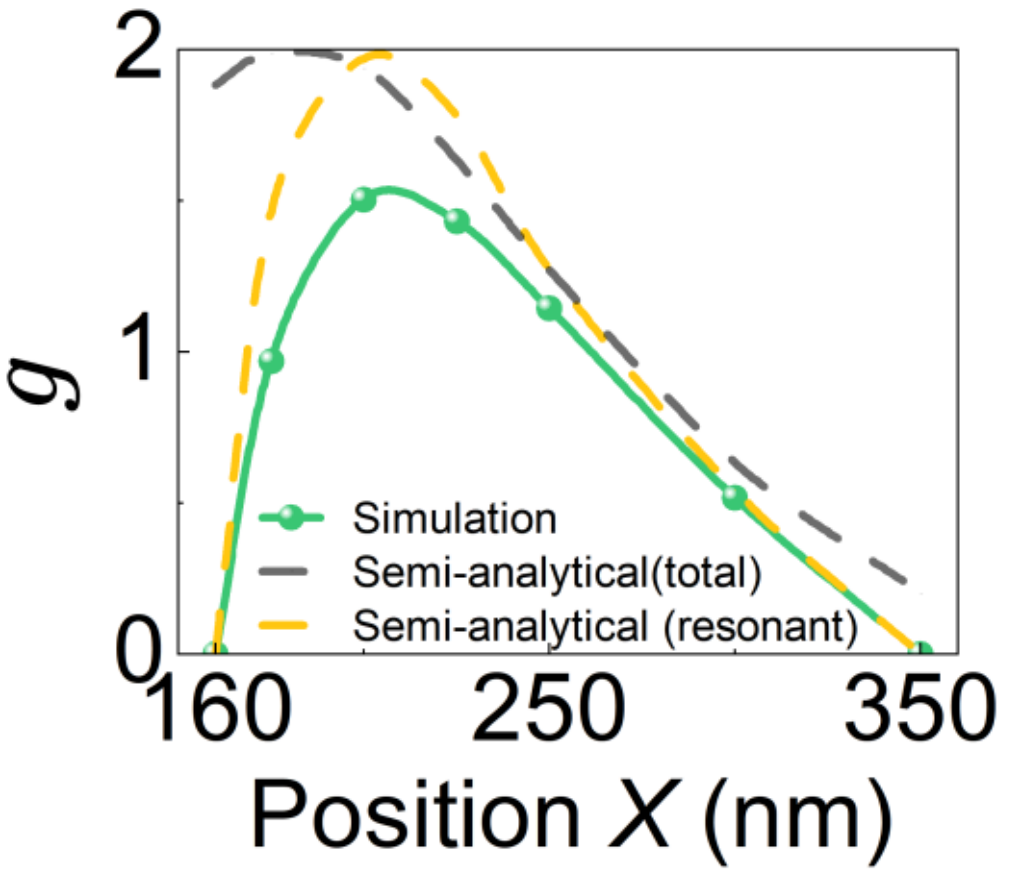

**Figure S5.** $g$-factor versus emitter position $x$ at $\theta$=90∘ with a fixed amplitude ratio $X/Y$ = 1, corresponding to the configuration in Fig. 3c of the main text. The semi-analytical prediction based on the total single-dipole rates (dashed blue line, "Semi-analytical (total)") shows a visible deviation from the full-wave simulation (gray diamonds), especially near the position where the magnetic resonant response is minimal.

The improvement after background subtraction is particularly evident when examined as a function of the emitter position. Figure S5 compares the semi-analytical predictions based on the total and background-subtracted rates across the full range of $x$ for the configuration corresponding to Fig. 3c of the main text. The deviation of the total-rate prediction is most pronounced near the position where the magnetic resonant response is minimal, as the non-resonant background dominates $\Gamma_{MM}$ and inflates the geometric mean estimate. After background subtraction, the deviation is significantly reduced, confirming that the discrepancy originates from the non-resonant background rather than a failure of the geometric mean relation itself. We note that the background-subtracted prediction is not expected to match the full-wave simulation exactly at all positions, as the simulation inherently includes the non-resonant background while the background-subtracted prediction isolates the resonant contribution.

## Part IV: Convergence toward the theoretical limit with increasing refractive index

Figure S6 presents the normalized decay rate spectra $F_+$ and $F_-$, together with the corresponding $g$-factor, for three representative refractive indices: $n$=3.5, $n$=4, and $n$=5. As $n$ increases, the cavity modes become more spectrally isolated, and the resonant intensity of the dominant mode is significantly enhanced—the peak Purcell factor $F_+$ increases from ~90 at $n$=3.5 to ~340 at $n$=5. The stronger resonant response further suppresses the relative contribution of the non-resonant background and improves numerical robustness against discretization errors. Consequently, the single-mode approximation underlying the semi-analytical model becomes progressively more accurate, and the peak $g$-factor converges toward the theoretical limit of 2, reaching ~1.97 at $n$=5.

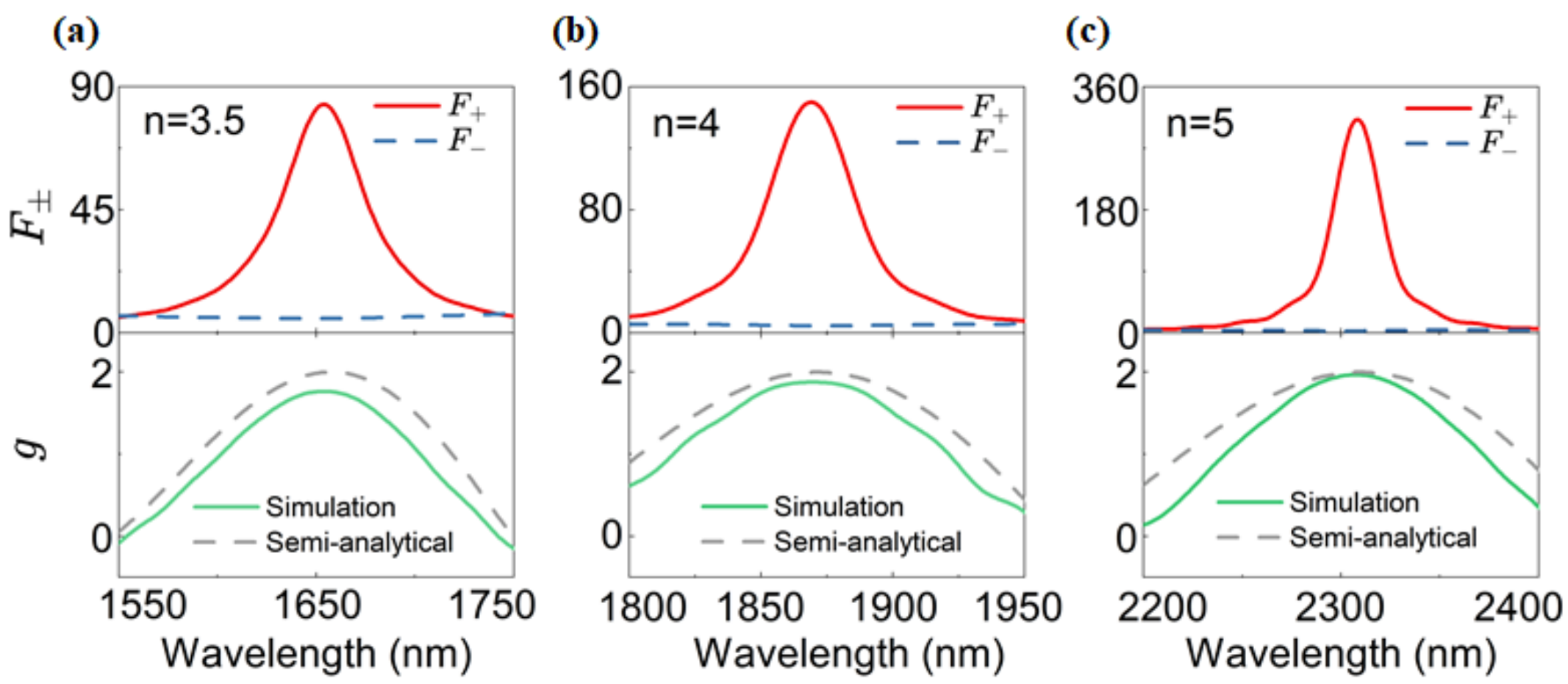

**Figure S6.** Refractive index dependence of the radiative asymmetry for the magnetic toroidal mode in the orthogonal configuration ($\theta$ = 90°). (a)–(c) Normalized decay rate spectra $F_+$ and $F_-$ for $n$=3.5, $n$=4, and $n$=5. (d)–(f) Corresponding $g$-factor from the full-wave simulation and the semi-analytical model. All other parameters are identical to those in Fig. 2 in the main text.

## Part V: Validation with and without vacuum voids

### Validation with dipoles embedded directly in the dielectric medium:

With the electric and magnetic dipoles embedded directly in the dielectric medium, the FDTD software reports apparent decay rates normalized to the dielectric background, which are smaller than the vacuum-normalized values by factors of $n$ for the electric dipole and $n^3$ for the magnetic dipole, following the standard scaling of dipole radiation power in a homogeneous dielectric medium. Without loss of generality, we set $X$=$Y$=1 and $\theta$=90° for simplicity. The software therefore apparent Purcell factors $A_1$, $B_1$ and $F_+^{app}$ that differ from the vacuum-normalized Purcell factors by $A$=$A_1n$, $B$=$B_1n^3$ and $F_+ = F_+^{app}(n+n^3)$. Here, $n$ is the refractive index of the dielectric, and both $F_+^{app}$ and $F_+$ denote Purcell factors of the enhanced (constructive-interference) parity-conjugated state—the former normalized to the dielectric background, the latter to vacuum. The vacuum-normalized cross-coupling rate $\Gamma_{EM}$ is retrieved from these apparent values via $2|\Gamma_{EM}| = F_+^{app}(n+n^3) - A_1n - B_1n^3$. Figure S7 compares the raw apparent spectra, the restored vacuum-normalized spectra, and the spectra obtained with vacuum voids. The restored $\Gamma_{EM}$ obtained from the above retrieval formula agrees well with the geometric mean prediction $\sqrt{\Gamma_{\mathrm{EE}}\Gamma_{\mathrm{MM}}}=\sqrt{A_1B_1}n^2$, confirming that the geometric mean relation holds for each situation.

**Influence of the vacuum voids on the Purcell factors:**

We note that the electric-dipole Purcell factor with the vacuum voids is enhanced by approximately a factor of two relative to the simple dielectric normalization prediction (i.e., $A \approx 2A_1n$ rather than $A = A_1n$, Figs. S7), while the magnetic-dipole factor remains close to the expected scaling ($B \approx B_1n^3$). The physical origin of this additional enhancement likely involves local field effects at the dielectric–void interface, although a complete understanding remains beyond the scope of the present work. For larger voids, both electric and magnetic Purcell factors tend to decrease moderately. Despite this quantitative shift in the specific values of $A$ and $B$, the theoretical framework of Eqs. (1)–(5) in the main text is entirely unaffected. All functional forms—including $g$=2sin$\theta$, the existence of an optimal matching condition, and the overall tunability of the radiative asymmetry—remain unchanged.

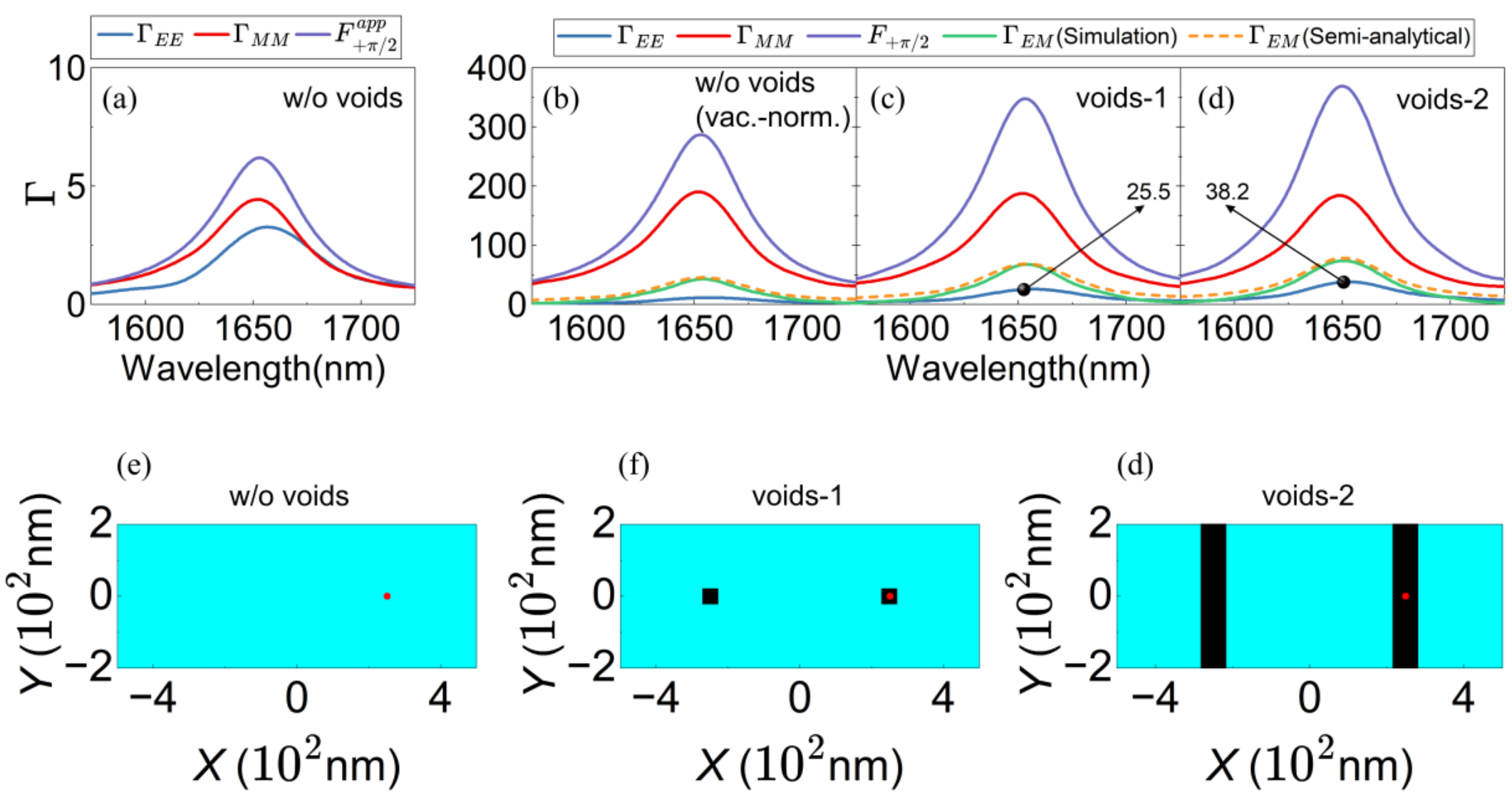

**Figure S7.** Validation of the geometric mean relation without vacuum voids. The other parameters are the same as that in Fig. 2 in the main text. (a) Raw apparent spectra directly output by the FDTD software with dipoles embedded directly in the dielectric medium: $A_1$, $B_1$ and $F_{+}^{app}$, all normalized to the dielectric background. (b) Vacuum-normalized spectra restored from the apparent values: $A$, $B$, $F_+$, $\Gamma_{EM}$ retrieved from simulation via $2\Gamma_{EM} = F_+^{app}(n + n^3) - A_1 n - B_1 n^3$, and the semi-analytical prediction $\sqrt{A_1 B_1} n^2$. (c) Spectra obtained with void-1, a cubic vacuum void of 30 nm side length (the configuration used in the main text). (d) Spectra obtained with void-2, a rectangular vacuum void $30 \times 400 \times 30$ $nm^3$. (e) Schematic of the w/o void configuration in the x*x*-y*y* plane, where the dipoles are embedded directly in the dielectric medium. (f) Schematics of the w/o void void-1 configuration in the *x*-*y* plane. (g) Schematic of the symmetric void-2 configuration in the *x*-*y* plane.

## Practical considerations for the choice of the void configuration

We note that the apparent values $A1$, $B1$, and $F_{+\pi/2}^{app}$ in the direct-embedding configuration are significantly reduced after normalization to the dielectric background, making them highly susceptible to non-resonant background interference. High-precision extraction of these values requires substantially finer computational grids and longer simulation times, while the residual numerical errors remain non-negligible. The introduction of the vacuum voids avoids this background normalization, yielding substantially larger apparent decay rates that are far more robust against numerical noise and enabling efficient parameter sweeps at modest computational cost. This practical consideration motivates the use of vacuum voids throughout the main text.

In all simulations with vacuum voids, the voids are introduced in parity-symmetric pairs: a void is placed at the emitter position, and an identical void is placed at the parity-symmetric counterpart location within the cavity. This symmetric placement rigorously preserves the overall parity symmetry of the cavity structure, ensuring that the voids themselves do not introduce any extrinsic parity asymmetry into the system. We note that the numerical results are insensitive to the presence of the second void. Nevertheless, the symmetric void configuration is adopted throughout as a matter of principle, to guarantee that any observed differentiation between parity-conjugated emitters originates solely from the coupling-induced parity breaking mechanism, rather than from any accidental asymmetry in the simulation setup.

## Part VI: Additional broadband responses

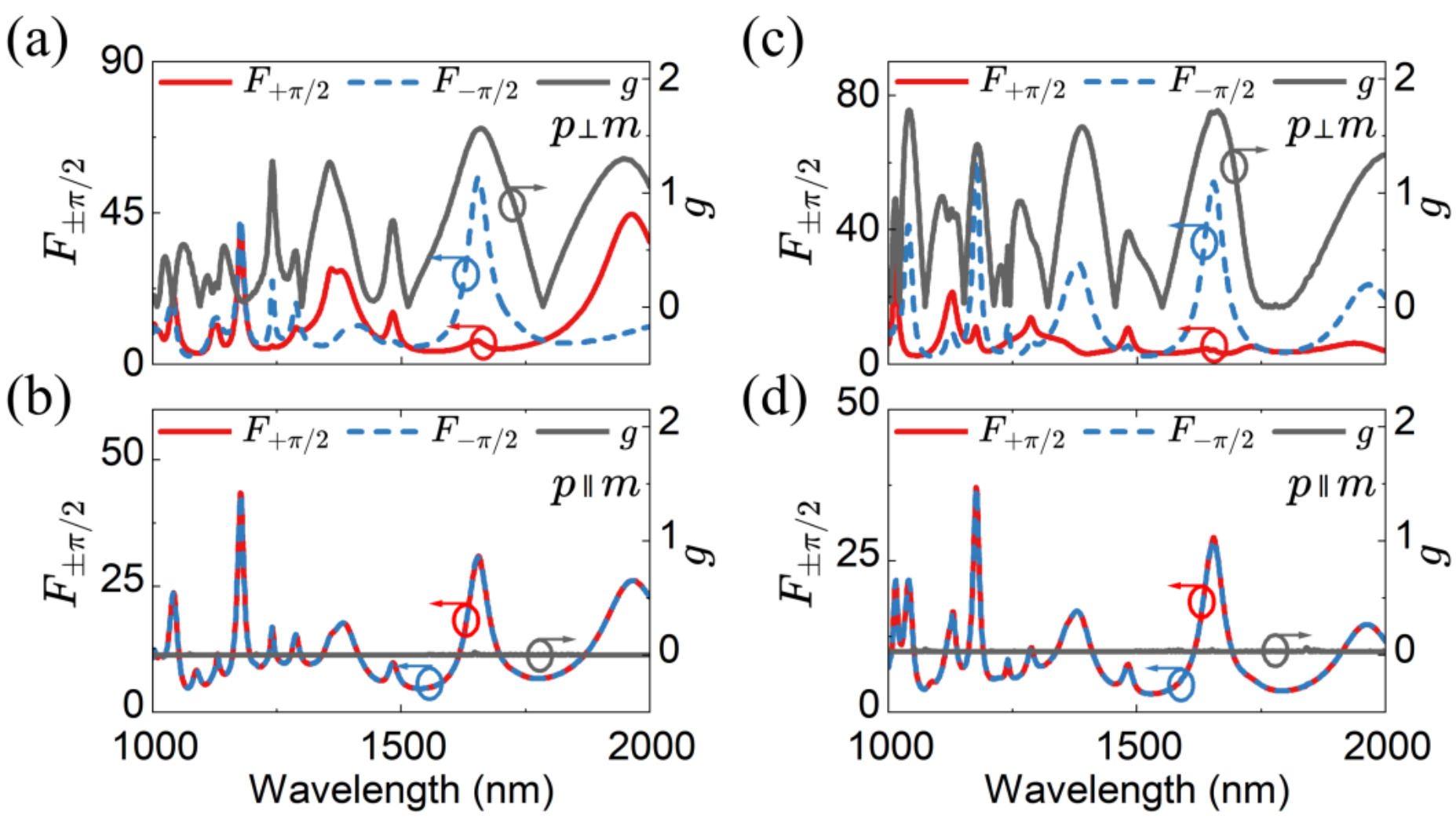

**Figure S8.** Broadband responses at two additional emitter positions for the magnetic toroidal mode. (a,b) Normalized decay rate spectra *F*+ and *F*− and the corresponding g*g*-factor at position 1 ($x$ = 100 nm, $y$=50 nm, $z$ = 0): (a) orthogonal configuration ($\theta$ = 90 °), (b) parallel configuration ($\theta$ = 0 °). (c), (d) Corresponding results at position 2 ($x$ = 450 nm, $y$=40 nm, $z$ = 0): (c) orthogonal configuration ($\theta$ = 90 °), (d) parallel configuration ($\theta$ = 0 °). All parameters are identical to those in Fig. 4 in the main text.